# Confidence regions for neutrino oscillation parameters from double-Chooz data

B. Vargas Perez, J. García-Ravelo, Dionisio Tun, and Jorge Garcia Bello

*Instituto Politécnico Nacional, Escuela Superior de Física y Matemáticas,
Departamento de Física, Unidad Profesional Adolfo López Mateos, C.P. 07738, Mexico City, Mexico*

Jesús Escamilla Roa

*Instituto Politécnico Nacional, Unidad Profesional Interdisciplinaria de Ingeniería Campus Hidalgo,
Ciudad del Conocimiento y la Cultura, C.P. 42162, San Agustín Tlaxiaca, Mexico
Universidad Autónoma del Estado de Hidalgo, Centro de Investigación en Matemáticas, Ciudad
Universitaria, C.P. 42184, Pachuca, Mexico*

In this work, an independent and detailed statistical analysis of the double-Chooz experiment is performed. In order to have a thorough understanding of the implications of the double-Chooz data on both oscillation parameters $\sin^2(2\theta_{13})$ and $\Delta m_{31}^2$, we decided to analyze the data corresponding to the Far detector, with no additional restriction. This differs from previous analyses, which only aim to estimate the mixing angle $\theta_{13}$, without mentioning the effects on $\Delta m_{31}^2$. By doing this, confidence regions and best fit values are obtained for $(\sin^2(2\theta_{13}), \Delta m_{31}^2)$. This analysis yields an out-of-order $\Delta m_{31}^2$ minimum, which has already been mentioned in previous works, and it is corrected with the inclusion of additional restrictions. With such restrictions it is obtained that $\sin^2(2\theta_{13}) = 0.084^{+0.030}_{-0.028}$ and $\Delta m_{31}^2 = 2.444^{+0.187}_{-0.215} \times 10^{-3}$ eV$^2$/c$^4$. Our analysis allows us to study the effects of the so-called "spectral bump" around 5 MeV; it is observed that a variation of this spectral bump may be able to move the $\Delta m_{31}^2$ best fit value, in such a way that $\Delta m_{31}^2$ takes the order of magnitude of the MINOS value. In other words, if we allow the variation of the spectral bump, then we may be able to determine both oscillation parameters using Far detector data only, with no further restrictions from other experiments. Finally, and with the intention of understanding the effects of the preliminary Near detector data, we performed two different analyses, aiming to eliminate the effects of the energy bump. As a consequence, it is found that unlike the Far detector analysis, the Near detector data may be able to fully determine both oscillation parameters by itself, resulting in $\sin^2(2\theta_{13}) = 0.095 \pm 0.053$ and $\Delta m_{31}^2 = 2.63^{+0.98}_{-1.15} \times 10^{-3}$ eV$^2$/c$^4$. The later analyses represent an improvement with respect to previous works, where additional constraints for $\Delta m_{31}^2$ were necessary.

## I. INTRODUCTION

The double-Chooz experiment estimated the reactor neutrino flux of the Chooz-B nuclear plant by means of its operating parameters. This flux, when interacting with the detector target, induces a number of inverse $\beta$ decays (IBD). This experiment was designed to run with two detectors located at $L_F \approx 1000$ m (Far), and $L_N \approx 400$ m (Near). But the current collaboration results report only far observations. The double-Chooz Near detector was finished in 2016, and only preliminary data have been published until now.

In particular, the double-Chooz Far detector reports fewer IBDs than those expected. If a neutrino oscillations model explains this deficit, then the oscillation parameters can be obtained from double-Chooz data.

In the simplified two-flavor oscillation model, the survival probability of a $\bar{\nu}_e$ with energy $E_\nu$(MeV) after traveling a distance $L$(m) is given as

$$P(\bar{\nu}_e \to \bar{\nu}_e) = 1 - \sin^2(2\theta_{13})\sin^2\left(1.27\frac{\Delta m_{31}^2 L}{E_\nu}\right). \quad (1)$$

The main objective of the double-Chooz experiment is the precise measurement of the mixing angle $\theta_{13}$ [1]. Table I shows some results for $\sin^2(2\theta_{13})$. In particular, the double-Chooz collaboration determined $\sin^2(2\theta_{13}) = 0.090^{+0.032}_{-0.029}$, without showing confidence regions and using the value obtained by MINOS of $2.44^{+0.09}_{-0.10} \times 10^{-3}$ eV$^2$/c$^4$ for $\Delta m_{31}^2$ [2].

TABLE I. Results of the different measurements made by the double-Chooz experiment. The first two (2012) include only reactor-on data and Gd captures. In 2013 the H-captures data were taken into account. From 2014 reactor-off data were included in addition to reactor-on data, only for Gd captures first and including H captures later. In the last one only Gd-captures data are taken into account.

|  | $\sin^2(2\theta_{13})$ |
|---|---|
| Matsubara [3] | $0.086 \pm 0.041^a \pm 0.030^b$ |
| Abe *et al.* [4] | $0.109 \pm 0.030^a \pm 0.025^b$ |
| Abe *et al.* [5] | $0.097 \pm 0.034^a \pm 0.034^b$ |
| Abe *et al.* [6] | $0.102 \pm 0.028^a \pm 0.033^b$ |
| Novella [7] | $0.102 \pm 0.043^c$ |
| Abe *et al.* [2] | $0.090^{+0.032}_{-0.029}{}^c$ |

[a]Statistical uncertainty.
[b]Systematic uncertainty.
[c]Total uncertainty.

In this paper we use the double-Chooz data in $\chi^2$ tests to determine the best fit for $\sin^2(2\theta_{13})$ and $\Delta m^2_{31}$ without assuming an *a priori* value for the last one to get both parameters as well as their confidence regions.

Consequently, double-Chooz data may be used for a unified analysis with other experiments where both oscillation parameters are obtained simultaneously by means of their corresponding data.

The organization of this work is as follows. In Sec. I the rate + shape (R + S) analysis is presented, aligned with Ref. [2]. The statistics defined there is used for Far detector analysis only. This section shows how the quantities that define the function $\chi^2_{R+S}$ are obtained. A consistent definition of the expected number of IBD is introduced in Sec. II. Both sections are used in Sec. III to estimate the oscillation parameters $\sin^2(2\theta_{13})$ and $\Delta m^2_{31}$ by minimizing the function $\chi^2_{R+S}$. Also the confidence regions for these parameters are obtained from a diagonal covariance matrix (DCM), and a full covariance matrix (FCM). The values obtained for $\Delta m^2_{31}$ are significantly different than those expected. In order to line up our results with the MINOS experiment, a $\chi^2_{R+S+M}$ test is presented.

In Sec. IV a discussion related to the spectral bump of the neutrino spectrum at 5 MeV is included to estimate their effect in the $\chi^2_{R+S}$ function.

Section V is devoted to the Far + Near detector data. With the purpose of determining the oscillation parameters $\Delta m^2_{31}$, $\sin^2(\theta_{13})$ and their confidence regions from double-Chooz data without using *a priori* the value of $\Delta m^2_{31}$ from another experiment, two convenient $\chi^2$ functions are introduced. These statistics also suppress the spectral bump effects mentioned before. To do this, we use the preliminary data from [8] as input to the formalism presented in the previous sections. The results obtained are promising and can be used when the collaboration releases new data.

Finally, our conclusions are given in Sec. VI.

## II. RATE + SHAPE ANALYSIS [2]

Neutrinos are detected through the positron kinetic energy, $E_{vis}$, in the energy range of 0.5 and 20 MeV, which is divided into 40 energy bins, accordingly to Table 15.2 in [9]. The rate + shape analysis is determined by the function

$$\chi^2_{R+S} = \sum_{i=1}^{40}\sum_{j=1}^{40} (N^{obs}_i - N^{exp}_i) M^{-1}_{ij} (N^{obs}_j - N^{exp}_j)$$
$$+ (\epsilon_a \epsilon_b \epsilon_c) \begin{pmatrix} \sigma_a^2 & \rho_{ab}\sigma_a\sigma_b & \rho_{ac}\sigma_a\sigma_c \\ \rho_{ab}\sigma_a\sigma_b & \sigma_b^2 & \rho_{bc}\sigma_b\sigma_c \\ \rho_{ac}\sigma_a\sigma_c & \rho_{bc}\sigma_b\sigma_c & \sigma_c^2 \end{pmatrix}^{-1} \begin{pmatrix} \epsilon_a \\ \epsilon_b \\ \epsilon_c \end{pmatrix}$$
$$+ \sum_{k=1}^{5} \frac{\epsilon_k^2}{\sigma_k^2} + 2\left(N^{obs}_{off} ln\left(\frac{N^{obs}_{off}}{N^{exp}_{off}}\right) + N^{exp}_{off} - N^{obs}_{off}\right). \quad (2)$$

In the first term, each energy bin requires $N^{obs}_i$ ($N^{exp}_i$), which is the observed (expected) number of IBD. A covariance matrix $M_{ij}$ is introduced to include the correlation terms among energy bins.

$N^{obs}_i$ were directly obtained from Fig. 21 in [2]. In this analysis we considered the 17 351 IBD candidates, $N^{tot}$, that occurred during 460.67 days, $T_{on}$. The first thirty-one readings are consistent with previous double-Chooz collaboration data [10].

$N^{exp}_i$ is proportional to the expected number of antineutrinos without oscillations, $n^{exp}_i$, and to the average survival probability in the $i$th energy bin, $\bar{P}_i$, closely related to the flavor-oscillation model (1). In the following section the explicit form of these quantities is given. For now we can write $N^{exp}_i$ as follows:

$$N^{exp}_i \sim n^{exp}_i \bar{P}_i(\theta_{13}, \Delta m^2_{31}). \quad (3)$$

The diagonal matrix elements $M_{ij}$ contain information on statistical and systematic uncertainties in each energy bin. The bin-to-bin correlations correspond to the off diagonal elements. $M_{ij}$ is discussed in more detail in [2] and is written as

$$M = M^{stat} + M^{flux} + M^{eff} + M^{Li/He(shape)} + M^{acc(stat)}. \quad (4)$$

The matrix elements $M_{ij}$ have been taken from Fig. 15.3 in [9]. In this figure there are five matrices that define the $M_{ij}$ numerical value. Since the information is presented through a color code, a basic software was necessary to decode it. We verified that the diagonal elements $M^{stat}_{ii}$, $M^{flux}_{ii}$, and $M^{eff}_{ii}$ previously published in [10] were found in the matrices of [9].

In addition to the uncertainties involved in the covariance matrix, eight systematic uncertainties are considered in the second and third terms of $\chi^2_{R+S}$, using $\epsilon_x$ parameters.

TABLE II. Uncertainties of the second term of Eq. (2). The numerical values were taken from [2].

| | |
|---|---|
| $\sigma_a$ | 0.006 MeV |
| $\sigma_b$ | 0.008 |
| $\sigma_c$ | 0.0006 MeV$^{-1}$ |
| $\rho_{ab}$ | −0.30 |
| $\rho_{bc}$ | −0.29 |
| $\rho_{ac}$ | $7.1 \times 10^{-3}$ |

Three of them are the coefficients of a polynomial for the visible energy variation,

$$\delta(E_{\text{vis}}) = \epsilon_a + \epsilon_b E_{\text{vis}} + \epsilon_c E_{\text{vis}}^2. \quad (5)$$

These are explicitly introduced in the second term of Eq. (2) by means of a matrix that contains uncertainties $\sigma_a$, $\sigma_b$, $\sigma_c$ and correlations $\rho_{ab}$, $\rho_{bc}$, and $\rho_{ac}$ (Table II).

The other sources of systematic uncertainties are considered through five parameters $\epsilon_k$, whose standard deviations are given in Table III.

The fourth term in (2) or $\chi^2_{\text{off}}$ is the contribution of two reactors off (2-off) data, in which $N^{\text{obs}}_{\text{off}}$ ($N^{\text{exp}}_{\text{off}}$) is the observed (expected) number of IBD candidates.

According to Ref. [2], $N^{\text{obs}}_{\text{off}} = 7$ and $N^{\text{exp}}_{\text{off}}$ is determined by the residual $\bar{\nu}_e$'s ($\epsilon_4$), and the background ($n^{\text{bg}}$), as

$$N^{\text{exp}}_{\text{off}} = \epsilon_4 \bar{P}_{\text{off}}(\theta_{13}, \Delta m^2_{31}) + n^{\text{bg}}, \quad (6)$$

where $n^{\text{bg}} = B \cdot T_{\text{off}}$.

$B = 1.56$ events/day is the total background rate provided by $T_{\text{off}} = 7.24$ days of reactor-off data, [9].

For IBD events with neutrons captured on Gd, $\bar{P}_{\text{off}}(\theta_{13}, \Delta m^2_{31})$ denotes the average survival probability on all the spectrum of antineutrinos with the reactor off, and is written as [9,11],

$$\bar{P}_{\text{off}}(\theta_{13}) = 1 - \sin^2(2\theta_{31}) \left\langle \sin^2\left(\frac{1.27 \Delta m^2_{31} \bar{L}}{E_\nu}\right) \right\rangle, \quad (7)$$

TABLE III. Miscellaneous uncertainties. Each one of them is related to its own pull parameter by means of the third term of Eq. (2). The pull parameters $\epsilon_1, ..., \epsilon_5$ are corrections to the predicted antineutrino spectrum. These are as follows: $\epsilon_1$, antineutrino spectrum error due to $\beta$ decays of $^8$He and $^9$Li; $\epsilon_2$, error due to $n + \mu$; $\epsilon_3$, accidentals; $\epsilon_4$, residuals; $\epsilon_5$, uncertainty of the squared mass differences $\Delta m^2_{31}$. The last one is removed later, [2].

| | |
|---|---|
| $\sigma_1$ | 0.13($d^{-1}$) |
| $\sigma_2$ | 0.038($d^{-1}$) |
| $\sigma_3$ | 0.0026($d^{-1}$) |
| $\sigma_4$ | 0.47(events) |
| $\sigma_5$ | $0.10 \times 10^{-3}$ eV$^2$ |

as suggested by Eq. (1). $\bar{L} = 1050\ m$ is the average distance from the Far detector to both reactors. The term in angle brackets results from averaging the survival probability (1) over all the energy range, $0.5\ \text{MeV} \leq E_{\text{vis}} \leq 20.0\ \text{MeV}$,

$$\left\langle \sin^2\left(\frac{1.27 \Delta m^2_{31} \bar{L}}{E_\nu}\right) \right\rangle$$
$$= \frac{1}{\Delta E_\nu} \int_{1.282\ \text{MeV}}^{20.782\ \text{MeV}} \sin^2\left(\frac{1.27 \Delta m^2_{31} \bar{L}}{E_\nu}\right) dE_\nu, \quad (8)$$

where

$$E_\nu \approx E_{\text{vis}} + \delta(E_{\text{vis}}) + 0.782\ \text{MeV} \quad (9)$$

is the energy of the incoming $\bar{\nu}_e$ written in terms of the corrected visible energy or positron kinetic energy, $E_{\text{vis}}$. The last term of (9) results directly from the observation of the IBD, and depends on the positron and nucleons masses, $m_{n^0} - m_{p^+} - m_{e^+} = 0.782\ \text{MeV}/c^2$.

## III. EXPECTED NUMBER OF IBD, $N^{\text{exp}}_i$

As introduced in Eq. (3), the expected number of IBD, $N^{\text{exp}}_i$, is closely related to the oscillation model and is defined as

$$N^{\text{exp}}_i = n^{\text{exp}}_i \left(1 + \frac{\epsilon_4 T_{\text{on}}}{T_{\text{off}} N^{\text{tot}}}\right) \bar{P}_i(\theta_{13}, \Delta m^2_{31})$$
$$+ \left(\epsilon_1 \frac{N^{Li+He}_i}{n^{\text{exp}}_i} + \epsilon_2 \frac{N^{n-\mu}_i}{n^{\text{exp}}_i} + \epsilon_3 \frac{N^{\text{acc}}_i}{n^{\text{exp}}_i}\right). \quad (10)$$

In this equation the expected neutrino spectrum without oscillations, $n^{\text{exp}}_i$, depends on the operating parameters of the nuclear reactor. These quantities are obtained from Fig. 21 at [2]; a discrepancy between the observed and the expected neutrino spectrum without oscillations between 4 and 6 MeV has been detected. This energy bump will impact the determination of the oscillation parameters. It is discussed in the next section.

The residual neutrinos $\epsilon_4$ are produced by the radioactive elements in the core of the nuclear reactors, even when they are turned off. The term

$$\epsilon_4 \frac{T_{\text{on}} n^{\text{exp}}_i}{T_{\text{off}} N^{\text{tot}}} \quad (11)$$

has been included to take into account this contribution to the spectrum.

Both types of events are influenced by the same average oscillation probability over each energy bin, $\bar{P}_i(\theta_{13}, \Delta m^2_{31})$.

Additionally, the three background sources, $\epsilon_1$, $\epsilon_2$, and $\epsilon_3$, mentioned in the description of Table III, are taken into account in Eq. (10).

Because we use the double-Chooz data to determine the best fit for $\sin^2(2\theta_{13})$ and $\Delta m_{31}^2$ without assuming an *a priori* value for $\Delta m_{31}^2$, the pull associated with the correction to this parameter, $\epsilon_5$, is not required anymore.

Therefore, the $\chi^2$ statistics (2) is a multiparametric function made of two oscillation parameters and seven pulls,

$$\chi^2_{R+S} = \chi^2_{R+S}(\theta_{13}, \Delta m_{31}^2, \epsilon_i, \epsilon_\alpha),$$
$$i = 1, \ldots, 4 \quad \alpha = a, b, c. \quad (12)$$

The $\Delta \chi^2_{R+S}$ function is defined as the difference between the $\chi^2_{R+S}$ function and its absolute minimum.

## IV. CONFIDENCE REGIONS OF OSCILLATION PARAMETERS $\theta_{13}$ AND $\Delta m_{31}^2$ WITH FAR DATA ONLY

We report the minimization of the function $\chi^2_{R+S}$ and its level curves considering the FCM $M$ in Fig. 1. Figure 2 takes into account only the diagonal elements of the covariance matrix $M$ (DCM). In both plots several local minimums are shown.

The absolute minimum (black star) in Fig. 1 has coordinates $(0.087, 27.043 \times 10^{-3} \text{ eV}^2/c^4)$ in the $\sin^2(2\theta_{13}) - \Delta m_{31}^2$ plane and the $\chi^2_{R+S}$ value of 37.17 for the FCM analysis. Another local minimum (white star) is $(0.090, 2.512 \times 10^{-3} \text{ eV}^2/c^4)$ and its $\chi^2_{R+S}$ value is 41.83. These points share close values for $\sin^2(2\theta_{13})$. Nevertheless their corresponding values for $\Delta m_{31}^2$ are significantly different. This effect is closely related to the existence of the energy bump of the neutrino spectrum. These points are listed in Table IV.

The absolute minimum obtained implies $\Delta m_{31}^2 = 27.043 \times 10^{-3} \text{ eV}^2/c^4$. This value is 1 order of magnitude higher than those reported elsewhere. However, given the quasiperiodic nature of the $\chi^2_{R+S}$ as a function of $\Delta m_{31}^2$, it can be argued that any one of the local minimums may be the right one, and then additional experimental data and/or improved models would be needed to discriminate between minimums. For this reason, in Sec. V we introduce data-data analyses that allow us the elimination of the energy bump effect. To do this, Near detector data are required.

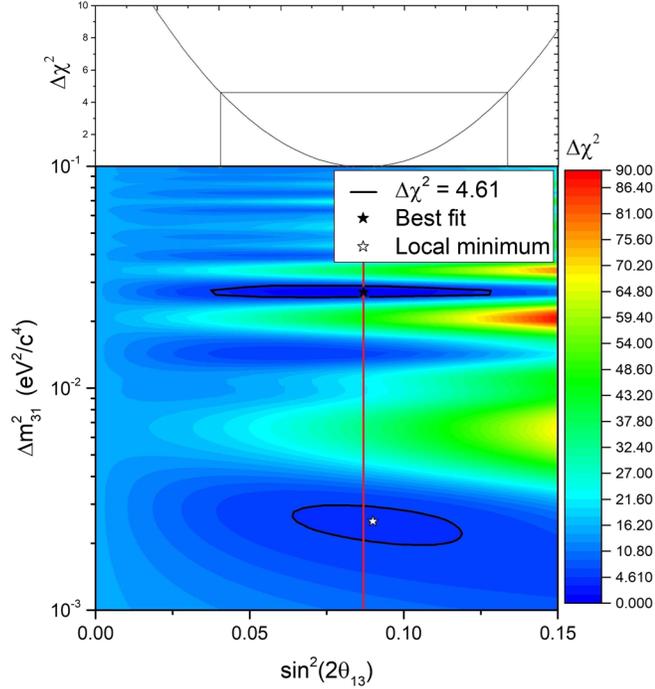

FIG. 1. Behavior of the $\chi^2_{R+S}$ statistics for FCM analysis. The confidence region up to 90% of C.L. for $(\sin^2 2\theta_{13}, \Delta m_{31}^2)$ is shown. Note that this confidence region is divided into two disjoint regions. The best fit is found at $\sin^2 2\theta_{13} = 0.087$, $\Delta m_{31}^2 = 27.043 \times 10^{-3} \text{ eV}^2/c^4$; this is an inconsistent $\Delta m_{31}^2$ value. Nevertheless, it is remarkable that a local minimum appears at $\sin^2 2\theta_{13} = 0.090$, $\Delta m_{31}^2 = 2.512 \times 10^{-3} \text{ eV}^2/c^4$; this minimum is included in the confidence region and is consistent with the $\Delta m_{31}^2$ value given by MINOS [12] (see Table IV). The plane $\sin^2 2\theta_{13} = 0.087$ is indicated with a red vertical line. The intersection of this plane with the $\chi^2_{R+S}$ level curves has been plotted in Fig. 4.

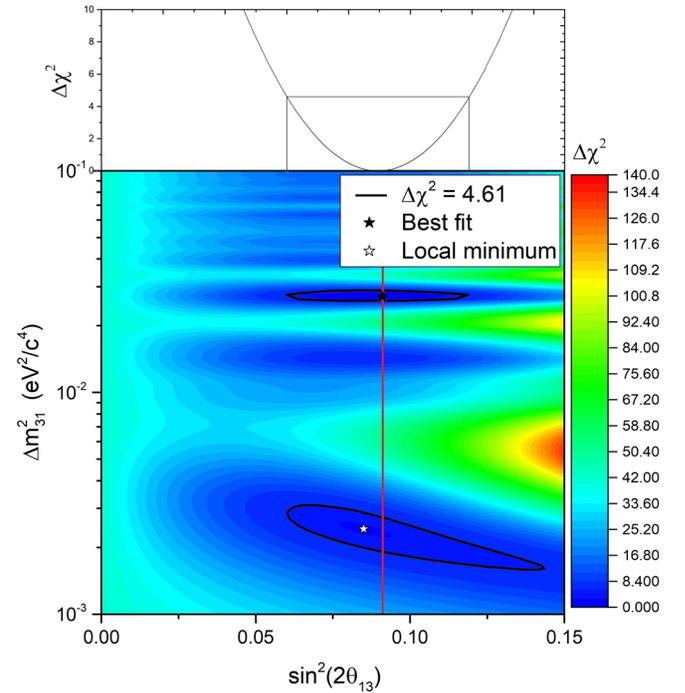

FIG. 2. Behavior of the $\chi^2_{R+S}$ statistics for DCM analysis. The confidence region up to 90% of C.L. for $(\sin^2 2\theta_{13}, \Delta m_{31}^2)$ shows two disjoint regions. The best fit is found at $\sin^2 2\theta_{13} = 0.091$, $\Delta m_{31}^2 = 27.043 \times 10^{-3} \text{ eV}^2/c^4$, again, an inconsistent $\Delta m_{31}^2$ value. This time, the local minimum appears at $\sin^2 2\theta_{13} = 0.085$, $\Delta m_{31}^2 = 2.422 \times 10^{-3} \text{ eV}^2/c^4$ and still belongs to the confidence region. This local minimum remains consistent with the $\Delta m_{31}^2$ value given by MINOS [12] (see Table IV). A discrimination criterion is needed. The plane $\sin^2 2\theta_{13} = 0.091$ is indicated with a red vertical line. The intersection of this plane with the $\chi^2_{R+S}$ level curves has been plotted in Fig. 4.

TABLE IV. Oscillation parameters found with the $\chi^2_{R+S}$ statistics for FCM and DCM analysis. The local minimums reported for the $\chi^2_{R+S}$ statistics are closer to those reported by double-Chooz [2], and MINOS [12] than the absolute minimums. In all cases, the $\Delta m^2_{31}$ units are $10^{-3}$ eV$^2$/c$^4$.

|  |  | $\chi^2_{R+S}$ | |
|---|---|---|---|
|  |  | FCM Fig. (1) | DCM Fig. (2) |
| Absolute minimum | $\chi^2_{m_3}$/D.O.F. | 37.17/39 | 40.07/39 |
|  | $\sin^2(2\theta_{13})$ | $0.087^{+0.047}_{-0.046}$ | $0.091^{+0.033}_{-0.029}$ |
|  | $\Delta m^2_{31}$ | $27.043^{+1.536}_{-1.217}$ | $27.043^{+1.456}_{-25.34}$ |
| First local minimum | $\chi^2_{m_1}$/D.O.F. | 41.83/39 | 43.34/39 |
|  | $\sin^2(2\theta_{13})$ | 0.090 | 0.085 |
|  | $\Delta m^2_{31}$ | 2.512 | 2.422 |
|  | Results of MINOS [12] | | |
| $\Delta m^2_{31}$ |  | $2.44^{+0.09}_{-0.10}$ (normal hierarchy) | |
|  |  | $2.38^{+0.09}_{-0.10}$ (inverted hierarchy) | |
|  | Results of double-Chooz [2] | | |
| $\sin^2(2\theta_{13})$ |  | $0.090^{+0.032}_{-0.029}$ | |
| $\chi^2_{min}$/D.O.F. |  | 52.2/40 | |

The simplest discrimination criterion is that which allows us to establish the value $\Delta m^2_{31}$, closest to that obtained in another experiment.

Consistently with [4], we introduce to the function (2) the additional term

$$\chi^2_{MINOS} \equiv \left(\frac{\Delta m^2_{31} - \Delta m^2_{MINOS}}{\sigma_{MINOS}}\right)^2, \quad (13)$$

where $\Delta m^2_{MINOS} = 2.44 \times 10^{-3}$ eV$^2$/c$^4$, and $\sigma_{MINOS}$ is the average of the $\Delta m^2_{31}$ uncertainties reported by MINOS, Table IV. So, by minimizing

$$\chi^2_{R+S+M} = \chi^2_{R+S} + \chi^2_{MINOS}, \quad (14)$$

TABLE V. Oscillation parameters found with the $\chi^2_{R+S+M}$ statistics for FCM and DCM analysis. The $\chi^2_{R+S+M}$ statistics change the $\chi^2_{R+S+M}$ local minimums to absolute ones, Fig. 3. The $\Delta m^2_{31}$ units are $10^{-3}$ eV$^2$/c$^4$.

|  |  | $\chi^2_{R+S+M}$ | |
|---|---|---|---|
|  |  | FCM | DCM |
| Absolute minimum | $\sin^2(2\theta_{13})$ | $0.092^{+0.058}_{-0.058}$ | $0.084^{+0.030}_{-0.028}$ |
|  | $\Delta m^2_{31}$ | $2.444^{+0.187}_{-0.194}$ | $2.444^{+0.187}_{-0.215}$ |
|  | $\chi^2_{min}$/D.O.F. | 41.85/40 | 43.32/40 |

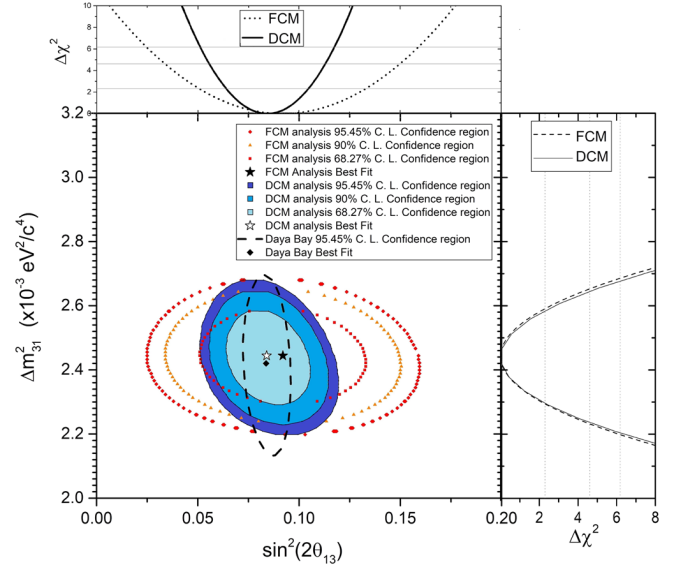

FIG. 3. Confidence regions up to 68.27%, 90%, and 95.45% for $\chi^2_{R+S+M}$, Eq. (14). As a consequence of the addition of $\chi^2_{MINOS}$ to the statistics, the absolute minimum is discarded. In this way, the best fit is found at $\sin^2 2\theta_{13} = 0.092$, $\Delta m^2_{31} = 2.444 \times 10^{-3}$ eV$^2$/c$^4$ for FCM analysis, and $\sin^2 2\theta_{13} = 0.084$, $\Delta m^2_{31} = 2.444 \times 10^{-3}$ eV$^2$/c$^4$ for DCM analysis. The wider region corresponds to the full analysis, and therefore, this one has greater uncertainties (see Table V). For comparison purposes we introduce the Daya Bay data for the parameters $\sin^2 2\theta_{13}$ and $\Delta m^2_{31}$ up to 95.45% of C.L. [13]. All the analyses are consistent to each other.

we obtain $\sin^2(2\theta_{13}) = 0.084^{+0.030}_{-0.028}$ and $\Delta m^2_{31} = 2.444^{+0.187}_{-0.215} \times 10^{-3}$ eV$^2$/c$^4$, with a $\chi^2_{R+S+M}$ minimum value given as $\chi^2_{min} = 43.32/40$ d.o.f. for DCM analysis. These results are presented in Table V.

The confidence regions generated from the $\chi^2_{R+S+M}$ statistics are consistent with those published in [9] and shown in Fig. 3.

## V. SPECTRAL BUMP EFFECTS

Figure 4 shows $\Delta\chi^2_{R+S}$ as a function of $\Delta m^2_{13}$ where $\sin^2(2\theta_{13})$ has been fixed at 0.087 and 0.091. These values correspond to the $\sin^2(2\theta_{13})$ coordinate of the absolute minimum obtained from the FCM and DCM analyses, respectively (Table IV). A succession of $\Delta\chi^2_{R+S}$ local minimums appear and are denoted as

$$\chi^2_{m_j} = \chi^2_{m_1}, \chi^2_{m_2}, \chi^2_{m_3}, \ldots \quad (15)$$

We can note that the $j$th minimum has the $\Delta m^2_{31}|_j$ coordinate; then

$$\chi^2_{m_j} = \Delta\chi^2_{R+S}(\Delta m^2_{31}|_j). \quad (16)$$

In the DCM analysis, the separation between two consecutive minimums of Fig. 4 is given as

$$\lambda_j = \Delta m^2_{31}|_{j+1} - \Delta m^2_{31}|_j; \qquad j = 1, 2, 3. \qquad (17)$$

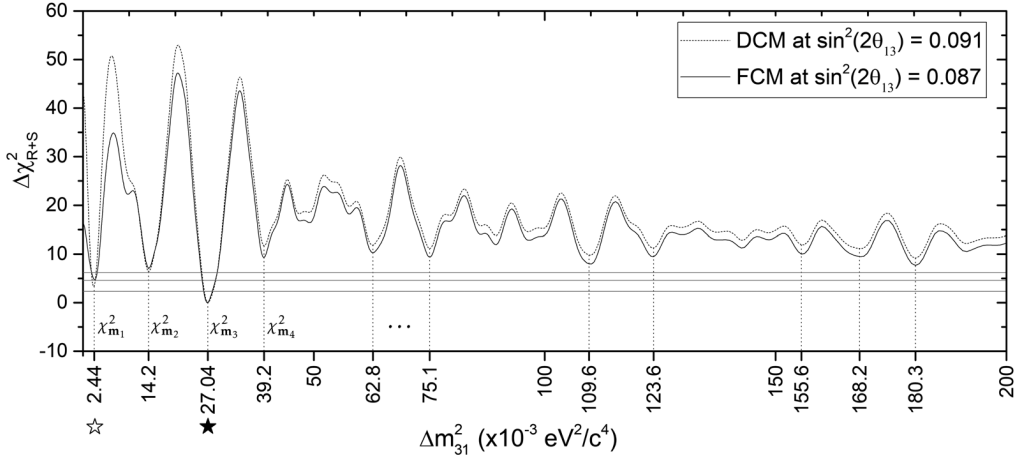

FIG. 4. $\Delta\chi^2_{R+S}$ profile as a function of $\Delta m^2_{31}$ for DCM analysis (dashed line) and FCM analysis (solid line), to their respective $\sin^2(2\theta_{13})$ best fit. Each analysis has several minimums. Notice that the jth minimum shares the $\Delta m^2_{31}$ coordinate for both analyses. The absolute minimum (black star) is found at an unacceptable value of $\Delta m^2_{31} = 27.043 \times 10^{-3}$ eV$^2$/c$^4$. The first minimum (white star) corresponds to $\Delta m^2_{31} = 2.3 \times 10^{-3}$ eV$^2$/c$^4$, which is close to the MINOS value $\Delta m^2_{31} = 2.44 \times 10^{-3}$ eV$^2$/c$^4$. The horizontal lines at $\Delta\chi^2_{R+S} = 2.3, 4.61,$ and $6.18$ represent the 68.27%, 90.0%, and 95.45% C.L., respectively. Notice that only the first local minimum and the absolute minimum fall into these regions.

Besides, we can identify the absolute minimum with $\chi^2_{m_3} = 0$ at $\Delta m^2_{31} = 27.043 \times 10^{-3}$ eV$^2$/c$^4$ and $\chi^2_{m_1} = 3.27$, with $\Delta m^2_{31} = 2.422 \times 10^{-3}$ eV$^2$/c$^4$, which is closer to the currently accepted $\Delta m^2_{31}$ value.

A weighted average of the neutrino energy can be defined as

$$\bar{E}_\nu = \frac{1}{40}\sum_{i=i}^{40} \omega_i E_{\nu,i} = 4.232 \text{ MeV}, \qquad (18)$$

where $\omega_i$ is the percentage of the observed IBD in each energy bin.

Substituting this value into the term $\sin^2(1.27\Delta m^2_{31}\bar{L}/\bar{E}_\nu)$, we found that it vanishes when $\Delta m^2_{31} = 0.012$ eV$^2$/c$^4$. This value is approximately equal to the average of $\lambda_1, \lambda_2,$ and $\lambda_3$. The small variation of these values may be attributed to the complex dependence of the $\chi^2_{R+S}$ on the squared sine function and to the average value of $\bar{E}_\nu$ used.

The Far detector results alone can be used to discuss how the spectral bump around 5 MeV in the neutrino spectrum affects the $\Delta m^2_{31}$ fit and how the distribution of $\chi^2_{m_j}$ might change.

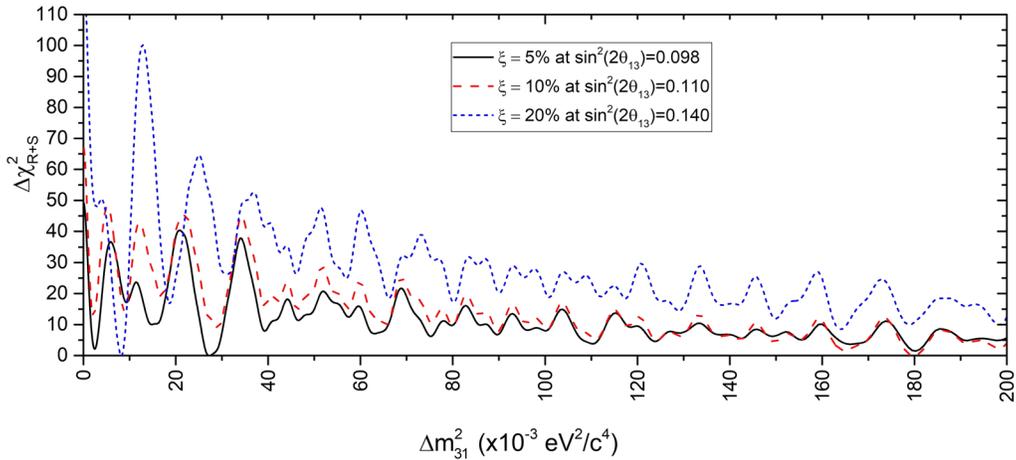

FIG. 5. $\Delta\chi^2_{R+S}$ profiles when a hypothetical source of rector neutrinos $\eta$ is added to the prediction in the energy bump, when the unknown source is 5% (black solid line), 10% (red dashed line), and 20% (blue short dashed line) of the total prediction and their respective $\sin^2(2\theta_{13})$ best fit. The oscillatory behavior of the $\chi^2$ functions remains, but the $\Delta m^2_{31}$ best fit adopts different values: $2.73 \times 10^{-2}$ eV$^2$/c$^4$ at $\xi = 5\%$, $1.79 \times 10^{-1}$ eV$^2$/c$^4$ at $\xi = 10\%$, and $8.20 \times 10^{-3}$ eV$^2$/c$^4$ at $\xi = 20\%$. The $\Delta m^2_{31}$ best fit is sensitive to the energy bump changes.

The origin of the energy bump is still undetermined, as discussed by [2,14]. The distortion seems to be hardly correlated to the reactor flux. This hypothesis was tested by the double-Chooz collaboration finding that the number of reactors on has influence on the distortion rate.

The pattern of minimums $\chi^2_{\mathbf{m}_j}$ is sensitive to the energy bump changes. As an example, we introduce a hypothetical source of rector neutrinos, given by $\eta_i \equiv \xi n_i^{\exp}$ added to the prediction in the energy bump. As a consequence, the oscillatory behavior of the $\chi^2$ functions remains, but the $\Delta m^2_{31}$ can diverge 2 order of magnitude higher than expected or even fall into the order set by MINOS Fig. 5. Note that when $\xi = 5\%$, the $\Delta m^2_{31}$ is still 1 order of magnitude higher than expected, and diverges 2 orders of magnitude when $\xi = 10\%$, but when $\xi = 20\%$ the difference of squared masses falls into $8.2 \times 10^{-3}$ eV$^2$/c$^4$.

Thus, the effect of the spectrum distortion is relevant but its source is unknown. If we want to obtain both parameters simultaneously, it is necessary to change our point of view, suppressing the energy bump as is discussed in the next section. This is encouraging to perform a unified analysis with other experiments.

## VI. CONFIDENCE REGIONS OF OSCILLATION PARAMETERS $\theta_{13}$ AND $\Delta m^2_{31}$ WITH FAR + NEAR DATA

Although the Near detector was built in 2016, only preliminary results have been published [8]. These preliminary double-Chooz two-detector results can be used as input to the formalism presented in Secs. II and III.

A direct comparison between two sets of data (a data-data analysis) has been considered to cancel the spectrum distortion for the determination of the oscillation parameters.

In order to perform a data-data analysis we are restricted to compare only the Far II to the Near data from [8]. In particular, we propose a $\chi^2_{(1)}$ statistics defined as

$$\chi^2_{(1)} = \sum_{i=1}^{40} \left( \frac{N^{\text{obs}}_{\text{Far},i} - \Omega_i N^{\text{obs}}_{\text{Near},i} \left( \frac{\bar{P}_i(\theta_{13}, \Delta m^2_{31}, L_{\text{Far}})}{\bar{P}_i(\theta_{13}, \Delta m^2_{31}, L_{\text{Near}})} \right)}{\sigma^{(1)}_i} \right)^2, \quad (19)$$

where $N^{\text{obs}}_{\text{Far/Near}}$ are the observed number of IBD candidates at the Far/Near detector in the bin with energy $E^{\text{vis}}_i$; $\Omega_i$ is a weight factor, and $\bar{P}_i(\theta_{13}, \Delta m^2_{31}, L_{\text{Far/Near}})$ is the averaged survival probability over each energy bin at the Far/Near detector.

This statistics suppresses the use of the prediction of the unoscillated reactor neutrino signal spectrum $n^{\exp}_{\text{Near/Far}}$ at the Near/Far detector.

Another way to define a data-data analysis is

$$\chi^2_{(2)} = \sum_{i=1}^{40} \left( \frac{\frac{N^{\text{obs}}_{\text{Far},i}}{\Omega_i N^{\text{obs}}_{\text{Near},i}} - \frac{N^{\exp}_{\text{Far},i}}{\Omega_i N^{\exp}_{\text{Near},i}}}{\sigma^{(2)}_i} \right)^2, \quad (20)$$

TABLE VI. Oscillation parameters found with the $\chi^2_{(1)}$ and $\chi^2_{(2)}$ statistics using Far II and Near data from [8]. The $\Delta m^2_{31}$ units are $10^{-3}$ eV$^2$/c$^4$. Notice in this case, the $\Delta m^2_{31}$ values obtained do not differ very much from those expected. The uncertainties are given by the 90% C.L. regions presented in Figs. 6 and 7. These results can be directly compared with those on Tables IV and V.

| | | $\chi^2_{(1)}$ | $\chi^2_{(2)}$ |
|---|---|---|---|
| Absolute minimum | $\chi^2$/D.O.F. | 53.4/40 | 42.1/40 |
| | $\sin^2(2\theta_{13})$ | $0.140^{+0.047}_{-0.043}$ | $0.095 \pm 0.053$ |
| | $\Delta m^2_{31}$ | $2.63^{+0.33}_{-0.55}$ | $2.63^{+0.98}_{-1.15}$ |

where $N^{\exp}_{\text{Far/Near},i} = n^{\exp}_{\text{Far/Near},i} \bar{P}_i(\theta_{13}, \Delta m^2_{31}, L_{\text{Far/Near}})$. This results from Eq. (10) when the background sources and the residual contribution are neglected.

The minimization of both data-data statistics leads to $\sin^2(2\theta_{13}) = 0.140^{+0.047}_{-0.043}$ and $\Delta m^2_{31} = 2.63^{+0.33}_{-0.55} \times 10^{-3}$ eV$^2$/c$^4$ for $\chi^2_{(1)}$, and for $\chi^2_{(2)}$, $\sin^2(2\theta_{13}) = 0.095 \pm 0.053$ and $\Delta m^2_{31} = 2.63^{+0.98}_{-1.15} \times 10^{-3}$ eV$^2$/c$^4$. These results are summarized in Table VI.

By means of the data-data analyses, the influence of the spectral distortion for the $\Delta m^2_{31}$ determination is highly suppressed. Figure 6 shows the 68.27%, 90%, and 95.45% C.L. regions. Three main points are remarkable.

(i) Data-data analyses no longer show two disjoint regions as the $\chi^2_{R+S}$ in Sec. III,

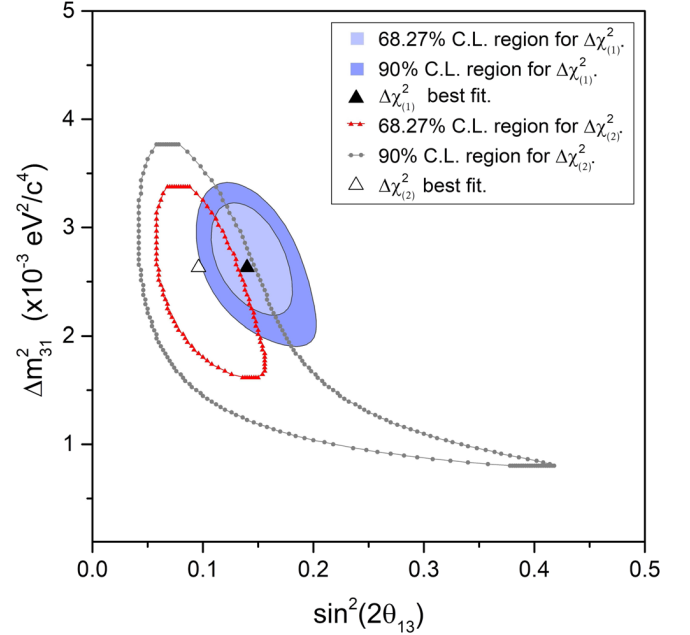

FIG. 6. 68.27%, 90%, and 95.45% C.L. regions for $\chi^2_{(1)}$ and $\chi^2_{(2)}$ and best fit. Through data-data analyses the spectral bump effect in the $\Delta m^2_{31}$ determination is highly suppressed. Even when the oscillatory behavior is still present, the $\Delta m^2_{31}$ is fully defined now; this is shown in Fig. 7.

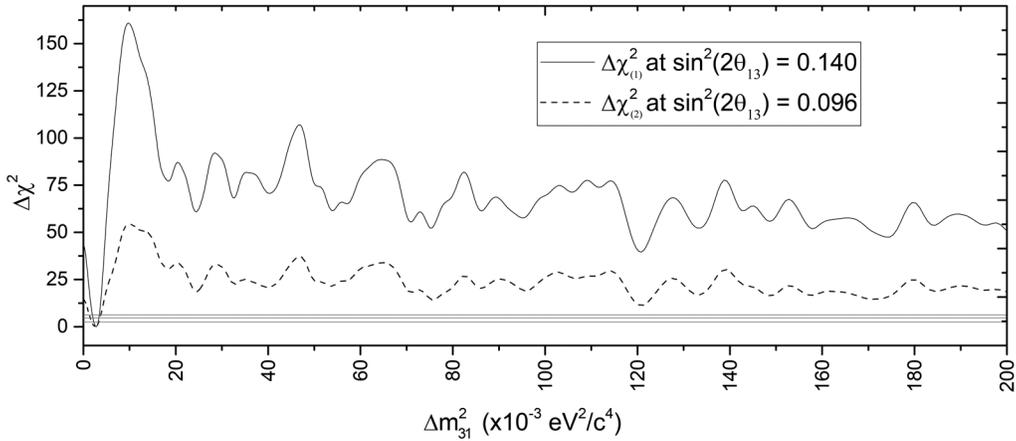

FIG. 7. $\Delta\chi^2_{(1)}$ (solid line) and $\Delta\chi^2_{(2)}$ (dashed line) profiles as a function of $\Delta m^2_{31}$ to their respective $\sin^2(2\theta_{13})$ best fit. The spectral bump effect in the $\Delta m^2_{31}$ determination is highly suppressed by means of data-data analyses. The oscillatory behavior is still present, but the $\Delta m^2_{31}$ is fully defined now and agrees with the currently accepted value for this parameter. The horizontal lines at $\Delta\chi^2_{R+S} = 2.3$, 4.61, and 6.18 represent the 68.27%, 90.0%, and 95.45% C.L., respectively. Notice that only the absolute minimum falls into these regions and it is near the expected one.

(ii) $\Delta m^2_{31}$ does not differ by 1 or more orders of magnitude with respect to MINOS,

(iii) $\chi^2_{(1)}$ and $\chi^2_{(2)}$ do not depend on external information, as $\chi^2_{R+S+M}$.

Even when both $\chi^2_{(1)}$ and $\chi^2_{(2)}$ statistics still have the oscillatory behavior in $\Delta m^2_{31}$, there is a well-defined difference between the absolute minimum and the local minimums, as can be seen in Fig. 7.

It is important to recall that in this section, preliminary data from [8] were used. In fact, the formalism described in Secs. II and III, in addition to the data-data statistics of this section, can be used to analyze the double-Chooz two-detector data to determine both $\sin^2(2\theta_{13})$ and $\Delta m^2_{31}$ without any restrictions from other experiments.

This work represents a useful tool to build a unified analysis of double-Chooz, Daya Bay, and RENO, as suggested in [15] and [16], even without solving the spectrum bump problem.

## VII. CONCLUSIONS

The proposed $\chi^2_{R+S}$ statistical analysis yields consistent results to those published by the double-Chooz collaboration using Far data. The approach followed allows us to generate the confidence regions for the oscillation parameters $\Delta m^2_{31}$ and $\sin^2(2\theta_{13})$ shown in Figs. 1 and 2.

The effect of the nondiagonal elements of the covariance matrix on the oscillation parameters can be compared with DCM analysis, which is in some sense a zero statistical error case based on the double-Chooz Far detector analysis only.

It is observed that in the FCM analysis the confidence regions are wider in $\sin^2(2\theta_{31})$, and therefore have greater uncertainties.

Each one of the FCM and DCM analyses reports the existence of a $\chi^2_{R+S}$ absolute minimum, corresponding to a $\Delta m^2_{31}$ value, which is inconsistent with the MINOS $\Delta m^2_{31}$ value. In fact, the double-Chooz Far data do not provide by themselves enough evidence to perform a squared mass difference $\Delta m^2_{31}$ estimation with the data available before 2016. However the first local minimum agrees with MINOS $\Delta m^2_{31}$ value, as shown in Fig. 4.

In order to force the first local minimum to become the absolute minimum we introduced the additional term (13), in $\chi^2_{R+S}$. Hence by minimizing the $\chi^2_{R+S+M}$ we got the best fit parameters, $\sin^2(2\theta_{13}) = 0.084^{+0.030}_{-0.028}$ and $\Delta m^2_{31} = 2.444^{+0.187}_{-0.215} \times 10^{-3}$ eV$^2/c^4$, as can be seen in Table V.

In Fig. 3 we have established the confidence regions for neutrino oscillation parameters $\theta_{13}$ and $\Delta m^2_{31}$ from double-Chooz Far data.

In Sec. IV we have introduced a hypothetical source of rector neutrinos to show how the spectrum distortion affects the oscillatory behavior of the $\chi^2$ functions and the $\Delta m^2_{31}$ value. We found that a correction of 20% in the expected spectrum distortion, independently of its source, corrects the order of magnitude of $\Delta m^2_{31}$, as indicated in Fig. 5.

To cancel the spectrum distortion in the determination of the oscillation parameters, we performed two data-data analyses, using preliminary two-detector data. In both cases, the $\Delta m^2_{31}$ values obtained are not so different than those currently accepted by the community as shown in Figs. 6 and 7 and Table VI.

Data-data analyses no longer show two disjoint regions as the $\chi^2_{R+S}$ in Sec. III. Also the value of $\Delta m^2_{31}$ found does not differ by 1 or more orders of magnitude with respect to MINOS and the whole analysis is independent of external information.

The formalism described in Secs. II and III in addition to the data-data statistics from Sec. V can be used to analyze the two-detector double-Chooz data to determine both $\sin^2(2\theta_{13})$ and $\Delta m^2_{31}$ without any restrictions from other experiments. In this way, this work extends the facilities of the double-Chooz experiment by allowing us to measure two oscillation parameters, $\Delta m^2_{31}$ and $\sin^2(2\theta_{13})$.

This work might contain elements of a future unified analysis with other experiments, such as Daya Bay and RENO even with the spectrum bump problem.

## ACKNOWLEDGMENTS

B. V. P. acknowledges the Escuela Superior de Física y Matemáticas, Instituto Politécnico Nacional, for the hospitality during his PhD studies in sciences. We also thank the kind referee for the positive and invaluable suggestions that have improved the manuscript greatly. Special thanks go to Karla Rosita Téllez Girón Flores for her suggestions. This work was partially supported by COFAA-IPN, Grants No. SIP20180062 and No. SIP20170031 IPN and the Consejo Nacional de Ciencia y Tecnología through the SNI-México.